\documentclass[useAMS,usenatbib,fleqn]{mn2e}
\usepackage{graphicx,amsmath,amssymb, aas_macros,times}
\usepackage[usenames,dvipsnames]{xcolor}
\usepackage{color}
\newcommand{\beq}{\begin{equation}}
\newcommand{\eeq}{\end{equation}}
\newcommand{\beqarray}{\begin{eqnarray}}
\newcommand{\eeqarray}{\end{eqnarray}}
\title{How Many of the Observed Neutrino Events Can Be Described by Cosmic Ray Interactions in the Milky Way?} 

\author[Joshi et al.]
{
Jagdish C. Joshi$^{1}$\thanks{jagdish@rri.res.in}, Walter Winter$^{2}$\thanks{winter@physik.uni-wuerzburg.de}, 
Nayantara Gupta$^{1}$\thanks{nayan@rri.res.in} \\
$^{1}$ Raman Research Institute, Sadashivanagar, Bangalore 560080, India\\
$^{2}$ Institut f\"ur theoretische Physik and Astrophysik, Universit\"at 
W\"urzburg, Am Hubland, D-97074 W\"urzburg, Germany \\
}

\begin{document}

\maketitle


\begin{abstract}
 Cosmic rays diffuse through the interstellar medium and interact with matter and radiations as long as they are trapped in the Galactic magnetic field. The
 IceCube experiment has detected some TeV-PeV neutrino events whose origin is yet unknown.We study if all or a fraction of these events can be described by
 the interactions of cosmic rays with matter. We consider the average target density needed to explain them for different halo sizes and shapes, 
 the effect of the chemical composition of the cosmic rays, the impact of the directional information of the neutrino events, and the constraints 
 from gamma ray bounds and their direction. We do not require knowledge of the cosmic ray escape time or injection for our approach.  
 We find that, given all constraints, at most 0.1 of the observed neutrino events in IceCube can be described by cosmic ray interactions with matter.
 In addition, we demonstrate that the currently established chemical composition of the cosmic rays contradicts a peak of the neutrino spectrum at PeV energies.
 
 \end{abstract}
\date{\today}
\maketitle
Cosmic ray propagation in our Galaxy has been studied in the past several 
decades using many models and with increasing complexities to explain the observational results successfully. The transport equation written by \cite{gs} contains 
various terms to include the possible gains and losses in the flux of cosmic rays. The simple leaky box model and its variants
 were widely used to explain the observed secondary to primary GeV cosmic ray flux ratios \citep{ss,cr}. Cosmic rays diffuse through the Galaxy, interact with 
 matter and background radiations producing secondary particles of lower atomic numbers $(Z)$.
 More complex models of cosmic ray propagation including the effects of energy dependent diffusion coefficient $(D)$ and re-acceleration
 were subsequently introduced by \cite{gw,gaisser,vb,lst}.

In the present work we consider the steady state flux of cosmic rays for the calculation of the diffuse neutrino flux produced in cosmic ray interactions,
directly based on cosmic ray observations. 
Thus our results neither depend on the unknown injection spectrum, nor on the escape time of very high energy cosmic rays.

The detection of very high energy and ultrahigh energy cosmic rays by air shower experiments \citep{ks, yk, agasa, hires, hv, auger} have an enormous impact on our
 understanding of the high energy phenomena in the universe. The compilation of cosmic ray data from various air-shower experiments show a knee region near 
 3~PeV and ankle region near $10^4$ PeV in the all particle cosmic ray spectrum \citep{gaissernew}.  

If we consider the propagation of these cosmic rays within the Galaxy, secondary gamma rays and neutrinos will be produced by their interactions
with Galactic matter~\citep{stecker, evoli, nayan1, nayan2}. 
 The IceCube experiment has detected some neutrino events in TeV-PeV energies 
which are unlikely to be of atmospheric origin \citep{ice1,ice2}. The implication of the IceCube neutrinos for cosmic ray transition models has been studied
in \cite{luis}, assuming that these could be of Galactic origin. Cosmic ray interactions in the inner Galaxy have been considered as the possible
origin of the some of the IceCube detected events and Fermi/LAT observed gamma rays in \cite{ner1}.
The five shower-like events correlated with the Galactic centre region \citep{soeb} could have originated from cosmic ray accelerations in SNR (supernova remnants). 
The correlation of the gamma ray and the neutrino fluxes and the Galactic origin of the IceCube events have been studied in \cite{ahlers}. They point out that within wide 
angular uncertainties off the Galactic plane, it is plausible that  about 10 events are of Galactic origin. 
Recently the sub-PeV and PeV neutrinos have been correlated with the cosmic rays above the second knee in the very high energy cosmic ray spectrum, for
various sources within hadronic interaction \citep{kohta}
and hypernova remnants have been suggested as their sources \citep{liu}.
The Neutrino events discovered by Ice-Cube can also come from p$\gamma$ interactions, as it is, for instance, discussed by \cite{winter,kohta1,stecker}.

In this work we study if the TeV-PeV neutrino events detected by IceCube above the atmospheric background have originated from interactions of very high 
energy cosmic rays (VHECRs) with Galactic matter or gas.
The interactions of cosmic rays with Galactic matter also lead to the production of high energy gamma rays which contribute to the background measured by Fermi/LAT. 

\section{Proton Interactions and Target Geometry} 

VHECRs interacting with Galactic matter give charged and neutral pions. 
The charged pions decay to muons and muon type neutrinos ($\pi^{\pm}\rightarrow \mu^{\pm}+\nu_{\mu} (\bar\nu_{\mu})$. The muons subsequently decay to electrons, 
electron type neutrinos and muon type neutrinos $(\mu^{\pm}\rightarrow e^{\pm}+\nu_{e}(\bar\nu_e)+\bar\nu_{\mu}(\nu_{\mu})$. The ratio of the neutrino fluxes of different
flavors produced in this way is $\nu_e+\bar\nu_e:\nu_{\mu}+\bar\nu_{\mu}:\nu_{\tau}+\bar\nu_{\tau}=1:2:0$.

The fluxes of neutrinos of each flavor are expected to be roughly equal on Earth 
after flavor mixing $\nu_e+\bar\nu_e:\nu_{\mu}+\bar\nu_{\mu}:\nu_{\tau}+\bar\nu_{\tau} \simeq 1:1:1$~\citep{gaisser}. For the numerical calculations, however, we compute the 
flavor mixing precisely using the current best-fit values from \cite{numix} (first octant solution).

For the description of the $pp$ interactions, we follow \cite{kelner}. 
The $pp$ interaction time is given by $t_{pp}(E_p)=1/(n_H\,\sigma_{pp}(E_p)\,c)$, where $n_H$ is the mean hydrogen number density of Galactic matter and the
cross section of the interaction is $\sigma_{pp}(E_p)=34.3+1.88 \, \mathrm{ln}(E_p/1 \, \mathrm{TeV})+0.25 \, ( \mathrm{ln}(E_p/ 1 \, \mathrm{TeV}))^2$~mb.  
The average (over different experiments) cosmic ray spectrum above 
100 TeV from \cite{gaissernew} has been approximated with power laws 
with several breaks for our calculation; the spectrum has been linearly interpolated among 
$(5,0)$, $(6.5,0)$, $(8.5,-0.85)$, $(9.7, -1.7)$, $(10.5,-1.7)$, $(11,-2.3)$ on a double log 
scale in $(\mathrm{log}_{10} E \, [\mathrm{GeV}],\mathrm{log}_{10} E^{2.6} J  \, \left[  \mathrm{GeV^{1.6} cm^{-2} \, s^{-1} \, sr^{-1} }  \right] )$.

The neutrino injection spectra $Q_\nu$ [$\mathrm{cm^{-3} \, s^{-1} \, GeV^{-1}}$] are given by
\begin{eqnarray}
Q_{\nu}(E_{\nu}) & = &  c \, n_H\, \int\limits_{0}^{1} \sigma_{pp}\left(\frac{E_\nu}{x} \right) \, N_p \left( \frac{E_\nu}{x} \right) \nonumber \\
& & \qquad \times f\left(x,\frac{E_\nu}{x}\right) \, \frac{dx}{x}  \label{nu_flux} 
\end{eqnarray}
for the appropriate flavor-dependent parameterizations of the distribution functions given in Eqs.~(62) and~(66) 
  of \cite{kelner}, which include the proper pion multiplicities. The integration over $x \equiv E_\nu/E_p$ is carried out to include the contributions 
  from all protons having energy equal to or higher than $E_{\nu}$. However, on the average $5\%$ of a proton's energy goes to a secondary neutrino,
  which means that the maximum contribution to the neutrino flux at energy $E_{\nu}$ comes from the protons of energy twenty times $E_{\nu}$. 
 Note that the neutrino injection is computed from the proper density $n_H$ [$\mathrm{cm^{-3}}$] and the steady state density $N_p$ [$\mathrm{cm^{-3} \, GeV^{-1}}$]
 obtained from solving the cosmic ray transport equation. 
 If we assume that the cosmic ray density is the same everywhere in the galaxy (or hydrogen halo), we can directly use the observed cosmic ray 
 flux to compute $N_p = 4 \pi/c \,  J_p$, where the fluxes are  given in units $[\mathrm{cm^{-2} \, s^{-1} \, sr^{-1} \, GeV^{-1} }]$. That is, 
 the neutrino production neither relies on the cosmic ray injection, nor on the cosmic ray escape time. The observed neutrino flux can be computed by 
  \begin{equation}
  J_\nu = \frac{1}{4 \pi} \, \int dV \frac{Q_\nu}{4 \pi r^2} \, ,
  \end{equation}
  where $r$ is the distance between Earth and production region. For a (hypothetical) spherical hydrogen halo with radius $R$ centred at Earth and a
  homogeneous target density, it is is easy to show that $J_\nu = Q_\nu \, R/(4 \pi) $. For an arbitrary halo shape, we can re-write  Eq.~(\ref{nu_flux}) as
\begin{eqnarray}
J_{\nu}(E_{\nu}) & = &  R_{\mathrm{eff}} \, n_H\, \int\limits_{0}^{1} \sigma_{pp}\left(\frac{E_\nu}{x} \right) \, J_p \left( \frac{E_\nu}{x} \right) \nonumber \\
& & \qquad \times f\left(x,\frac{E_\nu}{x}\right) \, \frac{dx}{x} \, .  \label{nu_flux2} 
\end{eqnarray}
Here the effective radius $R_{\mathrm{eff}}  \equiv   \int dV/(4 \pi r^2)$ for a homogeneous halo, integrated over the appropriate production region; for a halo
centered at Earth, one recovers $R=R_{\mathrm{eff}}$. If the hydrogen density or cosmic ray density depends on the location, this effect can be also expressed in
terms of the effective radius $R_{\mathrm{eff}}$ in a more complicated scheme; for a detailed study of the spatial
distribution of hydrogen and cosmic rays, see \cite{evoli} .

 In some models \citep{evoli} the average atomic hydrogen density in the Galaxy 
modelled with radii 10's of kpc and height 100's of pc calculated to be $\sim$ 0.5 $\mathrm{cm}^{-3}.$ The density of ionized, neutral and molecular hydrogen as a function of 
the height from the Galactic plane relative to the Earth's location and the radial distance from the Galactic centre have been calculated in \cite{feld} using the
gamma ray data observed by Fermi gamma ray space telescope.Relative to the Earth's location the density of atomic and molecular hydrogen gas drops from $1 \, \mathrm{cm}^{-3}$ to $0.1 \, \mathrm{cm}^{-3}$
within a distance of 1-1.5 kpc above the Galactic plane. The density of ionized hydrogen gas steeply falls from $0.3 \, \mathrm{cm}^{-3}$ to $0.001 \, \mathrm{cm}^{-3}$ within
the same distance. The hydrogen densities of $1 cm^{-3}$ are unlikely for the $10's$ kpc of spherical halo as discussed in \cite{dickey,kalb,blitz}.

It is expected to be much higher closer to the Galactic centre.
Please note that they have used a time dependent injection spectrum proportional to $E_p^{-2.4}$ and solved the diffusion equation to derive the steady state cosmic ray proton
spectrum. We are using the observed cosmic ray spectrum in our calculations.
We completely independently derive the average  hydrogen density from the neutrino observations, assuming that the observed events come from interactions between cosmic
rays and hydrogen within the halo. We consider different shapes of the hydrogen halo. The effective radii from Eq.~(\ref{nu_flux2}) for the different geometries and the 
Earth 8.33~kpc off the Galactic centre are listed in Table~\ref{xi}, where we denote the radius of the spherical region around the Galactic centre by $R_{\mathrm{GC}}$.
\begin{table}
\begin{center}
\begin{tabular}{lrrr}
\hline
 Shape & $R_{\mathrm{GC,kpc}}$ & $h_{\mathrm{kpc}}$ & $R_{\mathrm{eff,kpc}}$ \\
\hline
Spherical & 10. & & 7.2 \\
Spherical & 15. & & 13.3\\
Cylindrical & 10.  & 2.5 & 4.7 \\
Cylindrical & 10.  & 1.5 & 3.5 \\
Cylindrical & 10.  & 0.5 & 1.7 \\
Cylindrical & 10.  & 0.25 & 1.0 \\
Cylindrical & 15.  & 2.5 & 6.5 \\
Cylindrical & 15.  & 0.5 & 2.1 \\
Cylindrical & 15.  & 0.1 & 0.58 \\
\hline
\end{tabular}
\end{center}
\caption{\label{xi} The effective halo radius $R_{\mathrm{eff}}$, calculated for different halo shapes and parameters. Here $R_{\mathrm{GC}}$ refers to the radius around the
Galactic centre, and $\pm h_{\mathrm{kpc}}$ to the extension of the cylinder beyond the Galactic plane for the cylindrical shape.
}
\end{table}

In the following, we use $R_{\mathrm{eff}} = 10 \, \mathrm{kpc}$ or $R_{\mathrm{eff}} = 1 \, \mathrm{kpc}$ for different extreme models, but our results can be easily re-scaled with Table~\ref{xi}. 
While for the spherical halo around the Galactic centre and extending beyond Earth $R_{\mathrm{eff}} \sim 7 - 13 \, \mathrm{kpc}$ seems plausible, smaller values are 
obtained for the cylindrical halos: For realistic scale heights $h \lesssim 250 \, \mathrm{pc}$, $R_{\mathrm{eff}} \simeq 1 \, \mathrm{kpc}$.

\section{Effect of Cosmic Ray Composition}

The observed cosmic ray flux contains protons, helium, carbon, oxygen, iron and heavier nuclei. In \cite{gaissernew}, the helium nuclei 
flux exceeds the proton flux above 10 TeV and at 1 PeV helium and iron nuclei fluxes are  comparable (shown with curves of different colors in Fig.~4 of \cite{gaissernew}). 
At 100 PeV the cosmic ray flux contains mostly iron nuclei and at 1 EeV protons dominate over iron nuclei.  
 Each nucleon in the nucleus interact with a Galactic hydrogen atom and pions are produced which subsequently decay to neutrinos and gamma rays. In the case
of composite nucleus, the (observed) cosmic ray flux of nuclei with mass number $A$ is 
$J_A(E_A)=dN_A(E_A)/dE_A$.

We tested two different approaches to compute the neutrino flux for heavier compositions. One is essentially the superposition model: we assume that the
nucleus with mass number $A$ and energy $E_A$ behaves as $A$ nucleons with energy $E_A/A$. As a consequence, we can use Eq.~(\ref{nu_flux2}) to compute the neutrino
flux by replacing $J_p(E_p)=dN_p(E_p)/dE_p \rightarrow A^2 J_A(A E_p) = A^2 dN_A(A E_p)/dE_A$. For a simple power law with spectral index $\alpha$, one 
has $J_p(E_p) = A^{2 - \alpha} E_p^{-\alpha}$, and as a consequence, the result is identical to protons for $\alpha=2$. As another approach, we rather follow
\cite{anch} and take into account that the cross section  $\sigma_{Ap}$ is higher by a factor of $A^{3/4}$ than $\sigma_{pp}$. In this case, 
we can re-write Eq.~(\ref{nu_flux2}) as
\begin{eqnarray}
J_{\nu}(E_{\nu}) & = &  R_{\mathrm{eff}}  \, n_H\, \int\limits_{0}^{1} \sigma_{Ap}\left(\frac{E_\nu}{x_A} \right) \, J_A \left( \frac{E_\nu}{x_A} \right) \nonumber \\
& & \qquad \times A \, f\left(A x_A,\frac{E_\nu}{A x_A}\right) \, \frac{dx_A}{x_A} \, ,  \label{nu_flux2nucl} 
\label{nu_flux_nucleons}
\end{eqnarray}
where $x_A=x/A$ is the fraction of the nucleus' energy going into the neutrino. For a simple power law, this yields a neutrino flux $\propto A^{1.75-\alpha}$, which is about
a factor of $A^{0.25}$ smaller than the one of the superposition model, with some compensation by the slightly higher cross section. The reason is, roughly speaking, 
that the cross section of the nucleus is somewhat smaller than that of $A$ nucleons, because of the surface area/volume ratio $\sim A^{2/3}$. Note that these 
differences are very small (at the level of 20\%), and we use the (more realistic) model in Eq.~(\ref{nu_flux_nucleons}) in the following, which allows 
us to implement variable compositions easily.

\begin{figure}
\centering
\includegraphics[width=0.44\textwidth]{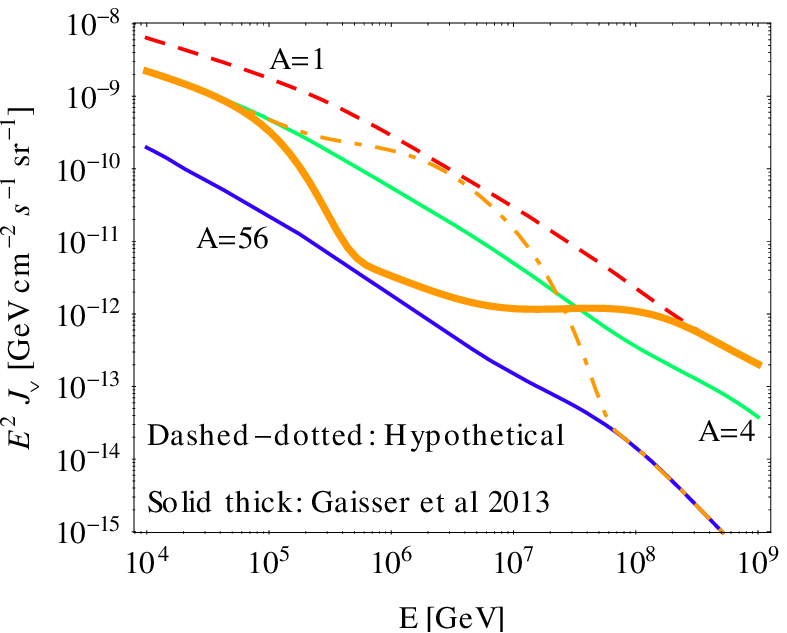}
\caption{Predicted neutrino flux for different cosmic ray compositions,
$n_H = 1 \mathrm{ cm^{-3}}$, and $R_{\mathrm{eff}}=1 \, 
\mathrm{kpc}$, corresponding to emission from a cylindrical halo with radius 10~kpc and half height 250~pc  ($\nu_\mu+\bar\nu_\mu$ flux including flavor mixing).} 
\label{pcomp}
\end{figure}

 Our predicted neutrino fluxes after flavor mixing for different cosmic ray compositions, $n_H = 1 \mathrm{ cm^{-3}}$, and $R_{\mathrm{eff}}=1 \, \mathrm{kpc}$ can be found in Fig.~\ref{pcomp}. 
 The Gaisser et al. composition has been adopted from Fig.~4 in \cite{gaissernew}, where we interpret $A(E_A)$ as a function of cosmic ray
 energy $E_A$ in Eq.~(\ref{nu_flux_nucleons}). In that case, we linearly 
 interpolate $A$ between $A=4$ at $5 \,\times 10^4 \, \mathrm{GeV}$, $A=4$ at $4 \,\times 10^6 \, \mathrm{GeV}$, $A=56$ at $8 \,\times 10^7 \, \mathrm{GeV}$, and $A=1$ at $10^9 \, \mathrm{GeV}$. 
 For the ``hypothetical model'', a helium composition between $5 \, \times 10^4 \, \mathrm{GeV}$ and $4 \, \times 10^6 \, \mathrm{GeV}$ has been chosen,
 then proton between $10^7$ and $10^8 \, \mathrm{GeV}$, and then iron at $10^9 \, \mathrm{GeV}$ (and higher), linearly interpolated among these values. 

First of all, since the flux roughly scales as $A^{2-\alpha}$, it is clear that the pure proton composition gives the highest flux and the pure iron
composition the lowest.  The Gaisser et al. model shows a iron composition at about $10^8 \, \mathrm{GeV}$, which leads to a dip in the neutrino flux at PeV energies,
exactly where the excess is observed. For comparison, we show a hypothetical model with a transition from heavier to lighter elements at these energies, 
with iron at the highest energies. This model produces a peak at exactly the right position, and therefore provides an especially good fit, but it contradicts  the iron knee in the cosmic ray composition observed by the KASCADE experiment~\citep{kampert}. Note that all cases with a composition heavier than hydrogen at 100~TeV lead to a predicted neutrino flux about one order of magnitude below the flux required to describe the IceCube observation~\cite{ice1}. 

We note that analytical estimates are not very accurate because a) the usual energy conservation arguments do not hold for spectra much steeper than $E^{-2}$, b) the cross 
section increases with energy which induces a small spectral tilt, and c) the distribution functions do have an impact.

\begin{figure*}
\includegraphics[width=0.35\textwidth]{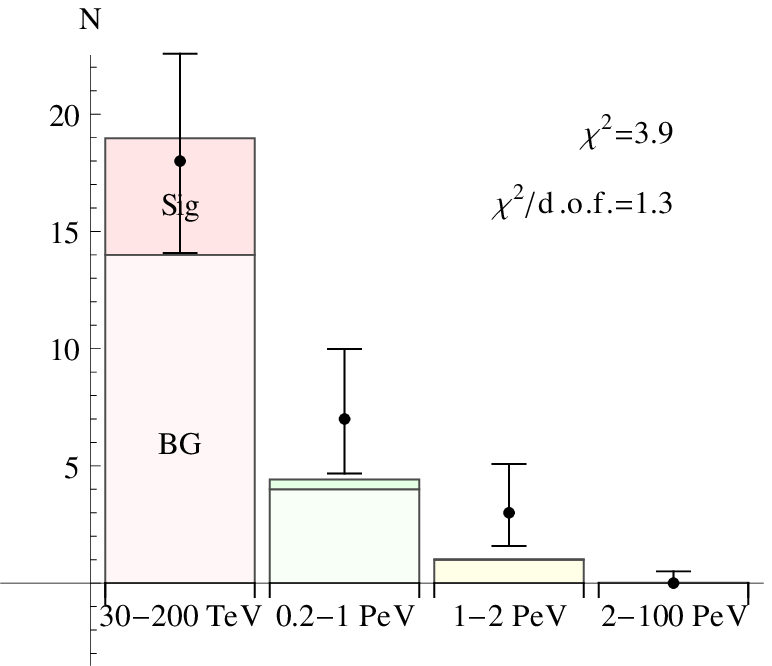}
\hspace{1.5cm}
\includegraphics[width=0.35\textwidth]{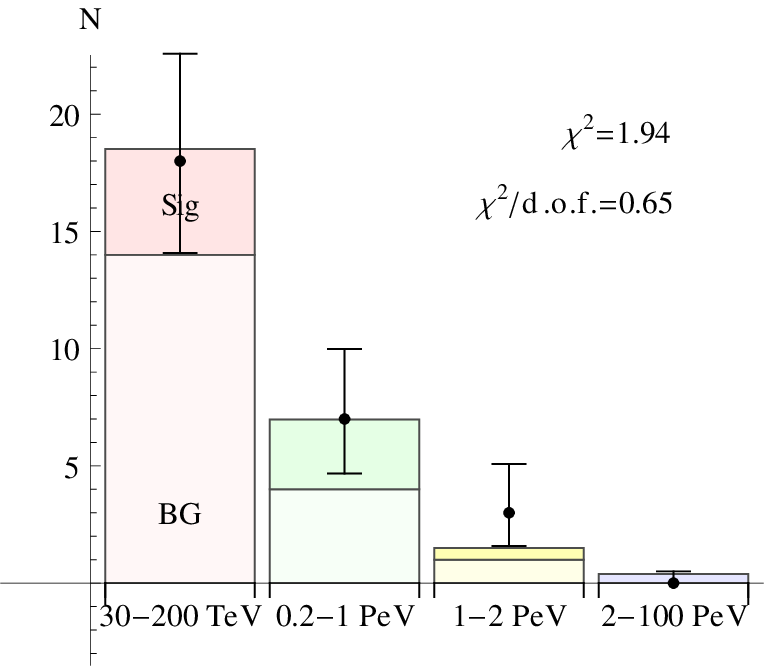} 
\caption{Observed (dots) and fitted (bars) event rates in the different energy bins for the Gaisser et al. 2013 and hypothetical models in the left and right panels, respectively. 
Here the model with directional information has been used. The required hydrogen densities are tabulated in Table 2.}
\label{fig:ppbestfit}
\end{figure*}

\section{Results for the target density}

The fluxes in Fig.~\ref{pcomp} depend on the product $R_{\mathrm{eff}} \times n_H$. Here we fit the computed neutrino spectra to the data in order to 
see what values can reproduce that, and what can be said about the fraction of neutrinos from cosmic ray interactions.  We follow the method described 
in \cite{winter} updated by~\cite{ice1}.
The neutrino events detected by the IceCube detector are  binned in four energy intervals 30-200 TeV, 0.2-1 PeV, 1-2 PeV and 2-100 PeV. We use
two different approaches: (1) Ignoring direction, we assume that all non-atmospheric events needs to be described by the interactions with hydrogen, 
computing the atmospheric background with the method in \cite{winter}; model ``All sky''. (2) We choose the events from the skymap~\cite{ice1} which
may potentially come from the cosmic ray interactions with the hydrogen halo within the directional uncertainties, and we correct for fraction of isotropically
distributed events which may fall into the Galactic plane; model ``Directional inf.''.\footnote{We remove the events at the lowest energies for that, as expected for 
the atmospheric background, in the ratio 2:1 showers to tracks. That is, the remaining signal events are 2, 3, 4, 13, 14, 15, 22, 25, 27 following the numbering in
Table~1 of~\cite{ice1}.} The rest of the events is treated as (extragalactic and atmospheric) isotropic background. In addition, we assume that the neutrino
directions are correlated with the diffuse gamma ray emission from the Galactic plane, which is limited to a Galactic latitude below $5^\circ$, see~\cite{FermiLAT:2012aa}. This
reduces the IceCube exposure to that flux by about a factor of ten because of the reduced solid angle.

\begin{table}
\begin{center}
\begin{tabular}{lrrrr}
\hline
 & \multicolumn{2}{c}{All sky} & \multicolumn{2}{c}{Directional inf.}  \\
 & \multicolumn{2}{c}{$R_{\mathrm{eff}}=10 \, \mathrm{kpc}$} & \multicolumn{2}{c}{$R_{\mathrm{eff}}=1 \, \mathrm{kpc}$}  \\
Composition & $n_H$ &  $\chi^2$ & $n_H$  & $\chi^2$  \\
 & $[ \mathrm{cm^{-3}} ]$ &  /d.o.f. & $[ \mathrm{cm^{-3}} ]$ & /d.o.f. \\
\hline
Hydrogen ($A=1$) & $1.6^{+0.3}_{-0.5}$ &  $1.9$ & $6.2^{+4.2}_{-3.7}$ &  $0.8$ \\
Helium   ($A=4$) & $5.9^{+1.7}_{-1.5}$ &  $2.1$ & $24^{+17}_{-15}$ &  $0.8$ \\
Iron     ($A=56$) & $130^{+38}_{-34}$   &  $2.5$ & $530^{+370}_{-330}$   &  $0.9$\\
\cite{gaissernew} &   $9.3^{+3.2}_{-2.8}$   & $5.1$ & $32^{+30}_{-26}$   & $1.3$\\
Hypothetical & $4.5^{+1.3}_{-1.2}$     &   $1.4$ & $20^{+13}_{-11}$     &   $0.7$ \\
\hline
\end{tabular}
\end{center}
\setcounter{table}{1}
\caption{\label{results} Best-fit hydrogen density for different cosmic ray compositions (first column) and two different composition and halo models. Here also the $1 \sigma$ errors from the fit to neutrino data are given, 
as well as the $\chi^2$ per degree of freedom for the fit.
The errors are non-Gaussian because of Poissonian statistics.}
\end{table}

We present our main results in Table~\ref{results}, where the best-fit target densities and the $\chi^2$/d.o.f. are shown for different composition models (rows), and two 
different extreme models for the directional information and halo sizes (columns). Note that $R_{\mathrm{eff}}=10 \, \mathrm{kpc}$ has been chosen for the ``all sky'' model, 
and $R_{\mathrm{eff}}=1 \, \mathrm{kpc}$ for the directional model; for different values, the results can be easily 
re-scaled using Table~\ref{xi}. From the all sky model, only the pure hydrogen composition produces realistic values for $n_H$, at the expense of a huge halo size.

For the model ``directional information'', the flux per solid angle in Eq.~(2) has to be divided by the solid angle assumed for the Galaxy ($0.087 \times 4\pi$) instead of $4 \pi$. Consequently, Figure~1 represents the solid angle-averaged flux. For the directional model, it is to be increased by the factor $1/0.087$ within the Galactic plane, and zero otherwise (c.f., Fig.~3, where the gamma-ray flux in the directional model is higher than in the all sky case). 

As a consequence, $n_H$ in Table~2 has to be lowered by this factor in the directional case, see updated Table~2.

Note that the statistics are good 
enough to derive lower bounds for the hydrogen density in the all sky case. In the directional model, the statistics are much poorer and the error bars therefore much larger.
Because of the small solid angle coverage of the signal, the required target densities are extremely large, which is unlikely. However, the event rates in IceCube from the direction of the 
Galactic plane can be well reproduced, see Fig.~\ref{fig:ppbestfit}. 
 For the Gaisser et al. 2013 cosmic ray composition (left panel), we obtain a relatively poor fit because of 
the dip at PeV (middle bins), exactly where the neutrino data require a peak (compare to Fig.~\ref{pcomp}). A better fit of the shape is, as expected, obtained for our hypothetical 
cosmic ray composition model, see right panel. Although this model is incompatible with cosmic ray composition data, it may serve as a proof of principle that one can produce a peak 
at PeV with composition changes only.  Note again that there is no direct dependence on the cosmic ray injection and escape time in our calculation.

\begin{figure}
\centering
\includegraphics[width=0.35\textwidth,angle=-90]{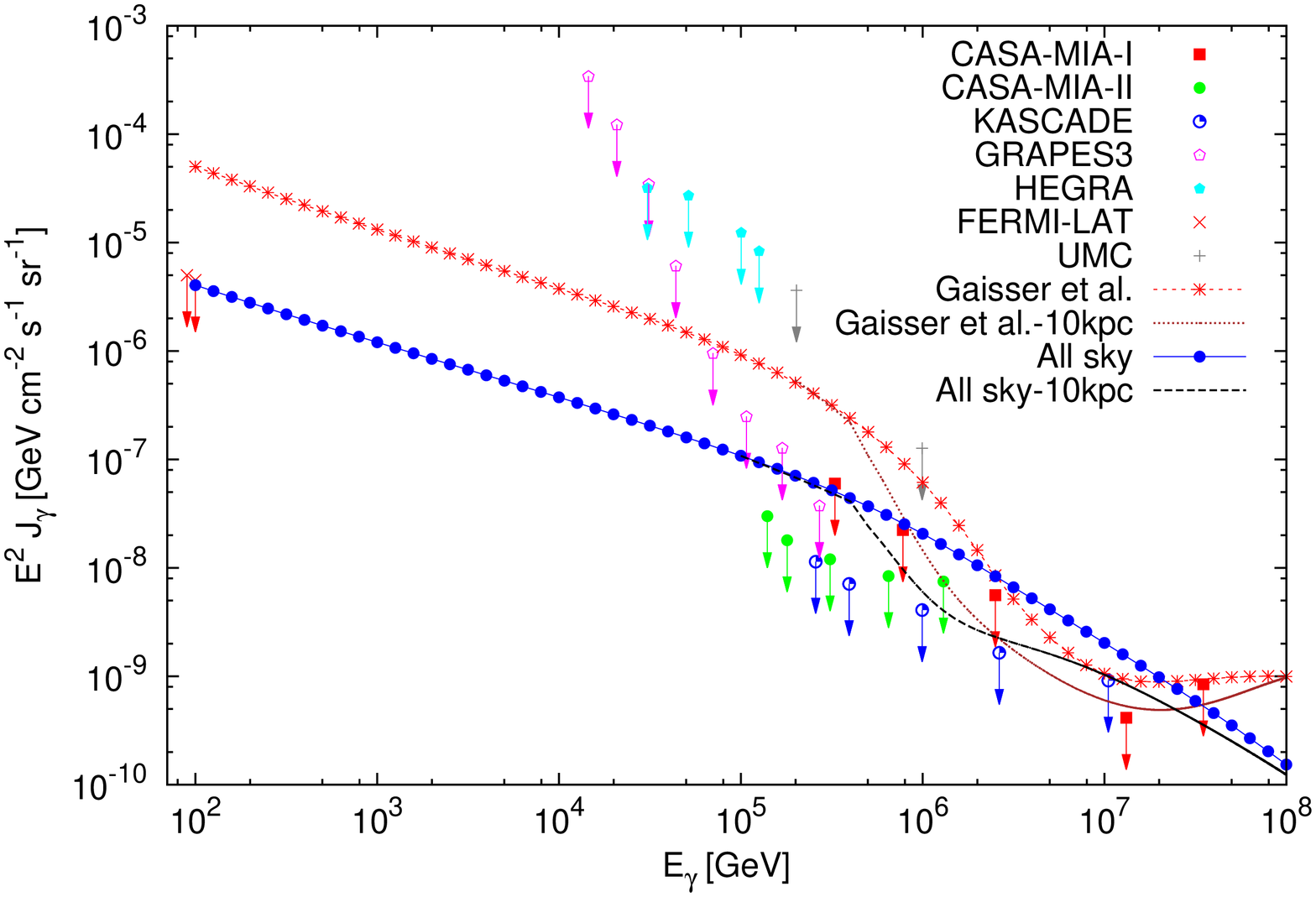}
\caption{Unattenuated gamma ray flux for two different models ($A=1$, All sky versus Gaisser et al. composition, directional information) compared with the limits from CASA-MIA-I \citep{casamia1}, KASCADE \citep{kascade},
HEGRA \citep{hegra}, GRAPES-3 \citep{grapes3} and UMC \citep{umc}. In addition, bounds from the Fermi-LAT Galactic plane diffuse emission \citep{FermiLAT:2012aa} (Fig. 17) 
and CASA-MIA~\citep{casamia2} are shown (CASA-MIA-II). The ``10 kpc'' curves show the effect of absorption due to the background radiation for 
a distance of 10~kpc~\citep{biermann}. The required hydrogen densities are tabulated in Table 2.}
\label{fig:gamma_flux}
\end{figure}

We have calculated the secondary very high energy gamma ray flux expected from $\pi^0$ decays produced in $pp$ interactions directly with \cite{kelner}. We show two different
cases in Fig.~\ref{fig:gamma_flux}: $A=1$ all sky model versus Gaisser et al. composition model with directional information.  Note that this result is shown for the best-fit 
of the models to neutrino data, i.e., the normalization is determined by the neutrino observation, and does not depend on $n_H$ or $R_{\mathrm{eff}}$ individually.
For illustration, we also show the curves for the gamma ray fluxes corrected for absorption due to the
background radiation with the mean free paths calculated in \cite{biermann} for $d=10 \, \mathrm{kpc}$. 
The upper limits on the diffuse gamma ray flux from various experiments are
 compared with our results. One strong constraint comes from the KASCADE and CASAMIA limits at a few hundred TeV. On the other hand, the Fermi-LAT observation
 at 100~GeV~\cite{FermiLAT:2012aa} does not impose a problem for the $A=1$ ``All sky'' model, whereas the directional model clearly exceeds the bound. The data above a
 few hundred TeV can be circumvented away by the attenuation of the gamma rays over long distances. The information given in Fig.~\ref{fig:gamma_flux} can be used to infer 
 the fraction of neutrino events which can come from the interactions in our Galaxy by rescaling the event rates to satisfy the bounds.

\section{Conclusions}

Taking into account the spectral shape of the observed neutrino spectrum, we have tested if it is plausible to describe the observed neutrino flux in the TeV-PeV range by 
interactions between cosmic rays and matter in the interstellar medium. We have discussed several composition models for the cosmic rays and several geometries for the
target matter halo. For the directional information on the neutrino events, we have chosen two possibilities: either all events above the atmospheric backgrounds are to be 
described by the matter interactions, or only the events compatible with the directions from the Galactic plane -- whereas the rest forms an isotropic (atmospheric and 
extragalactic) background. In the latter case, we have also taken into account a probable correlation with the diffuse gamma ray emission from the Galactic plane.

We have demonstrated that strong constraints arise from a) the expected target densities obtained from cosmic ray propagation models, b) bounds on the diffuse gamma ray emission 
from the Galactic plane, c) the measured cosmic ray composition contradicting the flux shape observed in IceCube, and d) the directional correlation with the 
diffuse gamma ray emission from the Galactic plane, limiting the expected solid angle of the signal flux. In the most plausible scenario (directional
information used, cosmic ray composition model by~\cite{gaissernew}), the required target density is about a factor of 100 above current expectations to
describe the neutrino events from the direction of the Galactic plane.In the Gaisser et al. composition model nine signal
events are obtained for the best-fit $n_H=32 \, \mathrm{cm}^{-3}$. In the directional case the average $n_H$ is $\sim 1 {cm}^{-3}$, about $9/32 \simeq 0.3$ events may come from 
cosmic ray interactions in the Milky Way.

Ignoring the directional information, a larger contribution $\simeq 1$ event is possible, 
taking into account the cosmic ray composition data, plausible halo sizes, and the gamma ray constraints -- which may serve as an upper limit for the estimate.
However, this scenario requires unrealistically large target densities.

 In conclusion, we have demonstrated that, taking into account the known constraints,
only a small fraction of the observed neutrino events may originate from the Galactic plane.

\section{Ackowledgment:}
NG acknowledges local hospitality during her visit to Universit\"at W\"urzburg.
WW acknowledges support from DFG grants WI 2639/3-1 and WI 2639/4-1, the FP7 Invisibles network (Marie Curie
Actions, PITN-GA-2011-289442), and the ``Helmholtz Alliance for Astroparticle Physics HAP'', 
funded by the Initiative and Networking fund of the Helmholtz association. We are grateful to M. Ahlers
for constructive comments.

\footnotesize{
\bibliography{ref}
}

\end{document}